\date{December 11, 2014}
\documentclass[12pt]{amsart}
\usepackage{latexsym,amsmath,amsfonts,amscd,amssymb}
\usepackage{graphics}
\newtheorem{theorem}{Theorem} %[subsection]

\theoremstyle{remark}

\newcommand{\cW}{{\mathcal W}}

\newcommand{\EE}{{\mathbb E}}

\newcommand{\RR}{{\mathbb R}}

\newcommand{\eps}{\epsilon}

\title{On Pareto theory of circulation of elites}


\author[R. P\'{e}rez Marco]{Ricardo P\'{e}rez Marco}
\address{CNRS, LAGA UMR 7539, Universit\'e Paris XIII,
99, Avenue J.-B. Cl\'ement, 93430-Villetaneuse, France}
\normalsize\email{ricardo.perez.marco@gmail.com}

\begin{document}
\begin{abstract}
We prove that Pareto theory of circulation of elites results from our wealth evolution model, Kelly criterion for optimal 
betting and Keynes' observation of ``animal spirits'' that drive the economy and cause that human financial decisions 
are prone to excess risk-taking.
\end{abstract}

\maketitle
%\noindent \emph{We dedicate this article to }

\section{Introduction.}

At the end of the XIXth century, Vilfredo Pareto in his economic and social studies made some fundamental discoveries. 
Based on his encyclopedic economic knowledge, his most famous observation is the universality and stability of the wealth distribution 
in different countries (Pareto law). In \cite{PM} we proposed a simple model of wealth evolution that 
implies Pareto distribution. Another observation by Pareto did set the foundation 
and motivation for his Theory of Elites. He observed that elites in different 
societies along history do renew systematically. This is called Pareto's "`Principle of Circulation of Elites"'.
He formulated a more elaborate theory of elites, but stated as an open problem the reason for the circulation of elites.
This is what he wrote about the distribution (``surface'' in his terms is the graphical representation of the histogram) of wealth distribution 
(see \cite{P2}, chapter VII, point 18, p. 386):

\medskip

\textit{``La surface nous donne una image de la soci\'et\'e. La forme extérieure varie peu, la partie int\'erieure est, au contraire, en 
perp\'etuel mouvement; tandis que certains individus montent dans les r\'egions sup\'erieures, d'autres en descendent. Ceux qui arrivent 
dans la zone la plus pauvre disparaissent; de ce cot\'e certains sont \'elimin\'es. Il est \'etrange, mais cel\`a est certain, 
que ce m\^eme ph\'enom\`ene se reproduit dans les r\'egions sup\'erieures. L'exp\'erience nous apprend que les aristocraties ne durent 
pas; les raisons du ph\'enom\`ene sont nombreuses et nous n'en connaissant que tr\`es peu, mais il n'y a aucun doute sur la 
r\'ealit\'e du ph\'enom\`ene.''}\footnote{\textit{``The histogram gives an image of the society. The external shape changes little, but 
the internal strucuture, at the opposite, is in perpetual motion. At the same time that some individuals climb into the upper ladder,
others fall down. Those arriving at the most poor region disappear and are eliminated. It is as strange as certain that the same 
phenomenon occurs in the upper regions. Experience tells us that the aristocraties do not last. The reasons are numerous but we know 
little, but there no doubt about the existence of the phenomenon.''}}

\medskip

A more elaborate ``Elite Theory'' is a central subject in Sociology and Political Science and was later developped by G. Mosca, R. Michels and others.
In these theories the ``Elite'' class is composed more about a rulling elite than about the wealthiest elite, but very often these two classes do coincide. 
In this article we only consider the elite class defined as the upper wealthy class. 

\medskip

\textbf{The explanation of Pareto's circulation of elites.}

\medskip

In the model of wealth evolution proposed in \cite{PM}, the individual wealth follows a series of financial decision modelled as a 
betting game: In each round a fixed proportion of 
wealth is wagered, which is lost or doubled with some probability $0<p<1$. The main theorem in \cite{PM} is the solution of the functional equation for 
an invariant distribution. This yields that the only invariant distributions obey Pareto law. We prove also that this distribution are stable 
for bounded $L^1$ perturbations. Having an invariant distribution does not mean that the individual wealth is constant. On the contrary, as already 
announced by the above citation by Pareto, the wealth of individuals is in constant evolution. ``en perp\'etuel mouvement'' in his own words.

Paradoxically, a positive expectation of wealth increase,
as one expects in a growing economy, does not result in an increase of the individual wealth. 
This is only true on average and individually only 
when the fraction of the wealth wagered 
is smaller than a fixed fraction $\gamma_0>0$ that depends on the probability $0<p<1$. 
This is a well known fact in gambling theory: The Kelly criterion 
gives the sharp fraction $0<\gamma^*<\gamma_0$ that one should risk in order 
to maximize the long term exponential growth, and a side result is that if we risk a fraction
$\gamma > \gamma_0$ then it is a mathematical theorem that 
we have almost sure ruin (with probability $1$). This means in the wealth evolution model that if we 
do have $\gamma > \gamma_0$ then as corollary we get Pareto's ``circulation 
of elite'', and more precisely we have that this occurs at all wealth levels, exactly as Pareto 
conjectured. Periodically individuals get wealthier 
and later on lose they fortune. 

\medskip

Now, the question reduces to why it is the rule that the fraction risked at each stage is superior 
to what Kelly criterion dictates. This cannot be a mathematical theorem but a common behavioural argument. 
It is a common observation that 
Kelly fraction is far more conservative than what the natural instincts dictate. Probability 
is not intuitive for the human mind, and we experience difficulties 
grasping the true nature of randomness.
Even for profesional gamblers it requires hard discipline to adjust to Kelly 
fraction. For example, a well known quotation by Jessy May, poker player and commentator, 
that applies to many other gambling or trading activities 
with advantage \cite{M}

\medskip

\textit{"Poker is a combination of luck and skill. People think mastering the skill 
is hard, but they're wrong. The trick to poker is mastering the luck.``}

\medskip

In economics and finance, a celebrated quotation of probabilist and economist J.M. Keynes 
argues into the same direction (\cite{K} p. 161):

\medskip

\textit{''Even apart from the instability due to speculation, there is the instability due to 
the characteristic of human nature that a large proportion of 
our positive activities depend on spontaneous optimism rather than mathematical expectations, 
whether moral or hedonistic or economic. 
Most, probably, of our decisions to do something positive, the full consequences of which will 
be drawn out over many days to come, can 
only be taken as the result of animal spirits—a spontaneous urge to action rather than inaction, 
and not as the outcome of a weighted 
average of quantitative benefits multiplied by quantitative probabilities.``}

\medskip

An attempt to explain these natural inclinations for risk-taking can be found is proposed by 
J. Coates, neuroscientist and former Wall Street trader, in his interesting 
book \cite{C} ''The hour between 
dog and wolf. Risk-taking, gut feelings and the biology of boom and bust`` where a biological 
origin is proposed for this type of behaviour. 
The point of view of J.M. Keynes is exposed very clearly (\cite{C} p.141):

\medskip

\textit{John Maynard Keynes, more than any other economist, understood these 
subterranean urges to explore, calling them ''animal spirits`` (...) He 
considered them the pulsing heart of the economy.(...) He suspected that business 
enterprise is no more driven by the calculation of odds than is an expedition to the 
South Pole. Enterprise is driven to a great extent by a pure love of risk-taking.''}

\medskip

The inadequacy of human mind to comprehend probability and randomness, and to calibrate 
risk properly, in particular hidden risks, permeates 
Nassim Taleb's ``risk philosophy'', and his ``black swans'' and ``antifragile'' theory. This was already 
masterly described in his first book ``Fooled by randomness: The Hidden Role of Chance 
in Life and in the Markets'' \cite{T}. 

\medskip

There is overwelming evidence that the risk-taking in financial decisions by average 
individuals go far beyond what Kelly criterion dictates, thus this 
explains Pareto's ``circulation of elite''.

\medskip

In the next sections we review the mathematical technical parts needed in our argument.

\section{Dynamical model of wealth evolution.}

A family of models of wealth evolution are presented in \cite{PM}, and 
for the simplest one a complete analysis is carried on. In order to explain the circulation of elites we need to use 
a more realistic model where the fraction
of the wealth at risk is totally lost in case of unfavorable outcome. 
For this model the Kelly criterion applies and this is 
key to our argument.

\subsection{Setup.}

Let $f(x)$ be the wealth distribution, i.e. $df= f(x) \  dx$ is the number of individuals with wealth 
in the infinitesimal interval $[x, x+dx[$. The 
distribution function 
$f : \RR_+ \to \RR_+$ is continuous, positive and decreasing and $\lim_{x\to +\infty} f(x) =0$. A distribution is of Pareto type 
if it presents a power law decay $x^{-\alpha}$ at $+\infty$, that is
$$
\lim_{x\to +\infty} -\frac{\log f(x)}{\log x} =\alpha >0 \ .
$$
The exponent $\alpha >0$ is \textit{Pareto exponent}. A distribution of the form $f(x) = C. x^{-\alpha}$ is called a Pareto distribution.
Smaller values of $\alpha$ indicate larger inequalities in wealth distribution. Notice that $\alpha >1$ is necessary for the distribution to 
be summable at $+\infty$, i.e. finite wealth at infinite (finitness near $0$ is not significant since the model aims to explain the tail behaviour at $+\infty$).

\subsection{Wealth dynamics.}

We focuss on the evolution of individual wealth. We assume that the evolution is based 
on two main factors:  Financial decisions, that we model 
as a betting game, and by public redistribution of wealth, that absorbs part of the individual wealth into public wealth.

For the first factor we model the financial decisions of each individual   
by a sequence of bets. Each financial decision turns out to be 
a bet, waging a proportion of his wealth. As a first approximation, we assume that the 
probability of success is the same for all agents and bets $0<p<1$ (this is the average probability). At each round, 
each agent risks the same percentage of his wealth, a fraction $\gamma >0$ (that is also an average). If 
he wins, his wealth is multiplied by the factor $1+\gamma$ and if he looses, with probability $q=1-p$, 
it is substracted the fraction of wealth 
put at risk, that is a percentage $\gamma$ of his wealth, that is, his wealth \footnote{This is 
what is different from the simplest model 
in \cite{PM} where after a loss, wealth was divided by $1+\gamma$. This allowed 
a complete computation of invariant distributions 
which is no longer the case here. The present model is more realistic and has the same assymptotic 
properties.} goes from $x$ to $x(1-\gamma)$.

Only considering this first factor, one round evolution the distribution transforms into the new distribution
\begin{equation} \label{eq_operator}
\cW (f) (x) = \frac{p}{1+\gamma} \ f \left ( \frac{x}{1+\gamma} \right ) + \frac{q}{1-\gamma } \ f \left ( \frac{x}{1-\gamma}\right ) \ .
\end{equation}
The operator $\cW$ is ``wealth preserving'', that is, in terms of $L^1$-norm we have
$$
||\cW (f)||_{L^1} = ||f||_{L^1} \ .
$$
The agents will only risk their capital if there is a positive expectation of gain. Giving a capital $x$, the expected gain at each round is 
$$
\EE(\Delta x) = p\gamma x -q \gamma x= (p-q) \gamma x = (2p-1) \gamma x \ ,
$$
thus we should assume that $p>1/2$.

There are other mechanisms that affect wealth evolution that we should consider, 
as for example inheritances that divide wealth, taxes, etc. 
As in \cite{PM} it is natural to consider a broader class of operators $\cW$ with a 
dissipative parameter $\kappa \geq 1$, the \textit{dissipative coefficient},
$$
\cW_\kappa (f) (x) = \frac{1}{\kappa} \cW (f) (x)=\frac{p}{\kappa (1+\gamma)}f ( x/(1+\gamma) ) + 
\frac{q}{\kappa (1-\gamma)} f( x/(1-\gamma)) \ .
$$
This operator is $L^1$ contracting for $\kappa >1$.

\subsection{Invariant distributions.}

Distributions invariant by the evolution operator $\cW_\kappa$  must satisfy 
the fixed point functional equation $\cW_\kappa (f)=f$, that is,
\begin{equation} \label{functional_eq}
f(x)= \frac{p}{\kappa (1+\gamma)}f \left ( x/(1+\gamma) \right ) + \frac{q}{\kappa (1-\gamma)} f(x/(1-\gamma))   \ .
\end{equation}

\subsection{Transformation of the functional equation.}

Considering the change of variables $F(x)=f(e^x)$, equation (\ref{functional_eq}) 
becomes a linear functional equation for $F:\RR \to \RR$
\begin{equation} \label{functional_eq}
a \ F(x+\lambda) - F(x) +b \ F(x-\mu) =0  \ ,
\end{equation}
where $\lambda = -\log (1-\gamma ) $, $\mu =\log (1+\gamma)$, $a=q/(\kappa (1-\gamma )) $ 
and $b=p/(\kappa(1+\gamma))$, with $a,b,\lambda ,\mu >0$.

We have a general theory of these type of functional equations. L. Schwartz (see \cite{S}, and also \cite{Ma}, \cite{Ka})
studied more general ``mean periodic'' smooth functions $F$ 
that satisfy a convolution functional equation of the form
\begin{equation} \label{distfunct_eq}
\omega \star F = 0 \ ,
\end{equation}
where $\omega$ is a compactly supported distribution. In our case, $\omega = a \ \delta_\lambda -\delta_0 + b \ \delta_{-\mu}$.
The way to study these equations 
is by Fourier transforming it (\`a la Carleman using hyperfunctions
in order to work in sufficient generality). Taking the Fourier transform of equation (\ref{distfunct_eq}), we get 
$$
\hat \omega . \hat F =0 \ .
$$
Thus if $\hat F$ is a Schwartz distributions then its support is contained in the set where $\hat \omega =0$, and if $\omega$ is 
a finite sum of atomic masses as in our model, this is the 
set of $\rho$'s which are roots of a Dirichlet polynomials. In our specific model we are led to solve 
the \textit{characteristic equation} in $\rho$:
\begin{equation}\label{char_eq}
a e^{\lambda \rho} -1 +be^{-\mu \rho} =0 \ .
\end{equation}

This equations yields particular solutions of the form $F(x)=e^{\rho x}$ (note that $\hat F=\delta_\rho$). 
In general the zeos of such a Dirichlet polynomial are 
contained in a vertical strip, thus the real part being bounded, and those with negative real parts give 
the admissible solutions all of them with Pareto assymptotics. 

The function $h(\rho)=a e^{\lambda \rho} -1 +be^{-\mu \rho}$ is convex since 
$h''(\rho) =a \lambda^2 e^{\lambda \rho} + b \mu^2 e^{-\mu \rho} >0$, and $h'$ has a unique 
zero at 
$$
\rho_0 = \frac{1}{\lambda+\mu} \log\left ( \frac{\mu b}{\lambda a}\right ) \ .
$$
In order to have admissible solutions with Pareto assymptotics at $+\infty$ converging to $0$, we need to have a negative solution 
to the characteristic equation. The condition $h(0)=a+b-1 <0$ is equivalent to the condition on $\kappa \geq 1$,
\begin{equation}\label{kappacondition}
\kappa > \kappa_0=\frac{1-(p-q) \gamma}{1-\gamma^2} \ .
\end{equation}
Thus we only have Pareto solutions with the right assymptotics if we have this condition on $\kappa$. It is interesting to 
note that $\kappa_0 >1$ is equivalent to $\gamma > p-q=2p-1 =\gamma^*$, and as we see in the next section, this is the Kelly condition.

\section{Kelly criterion.}

\subsection{Setup.}

We assume that we are playing a game with repetition. At each round we risk a fraction $0\leq \gamma \leq 1$ of our capital $X$. 
With probability $0 < p <1$ we win and the pay-off is the fraction of the capital we bet. This means that if $X$ is 
our current capital, if we lose the bet (which happens with probability
$q=1-p$) we substract $fX$ to our capital, and if we win, we add $fX$ to our capital . As we have already seen
the condition that the game has positive expectation is $p>1/2$.
If we play a game with advantage, then we expect an exponential growth 
of our initial bankroll $X_0$ if we follow a reasonable
betting strategy. We assume that there is no minimal 
unit of bet. By homogeneity of the problem, a sharp 
strategy must consist in betting a proportion $\gamma (p)$ of 
the total bankroll. Our bankroll after having played $n$ rounds of the 
game have is
$$
X_n=X_0 \prod_{i=1}^n (1+ \eps_i \gamma (p))
$$
where $\eps_i=+1$ if we won in the $i$-th round, and $\eps_i=-1$
if we lose in the $i$-th round. The exponential rate of growth of the bankroll is
$$
G_n =\frac1n \log \frac{X_n}{X_0}=
\frac1n \sum_{i=1}^n \log (1+\eps_i \gamma (p)).
$$
The Kelly criterion maximizes the expected value of the 
exponential rate of growth:

\begin{theorem} \textbf {(Kelly criterion)}
For a game with advantage, that is $p >1/2$, the expected value of the exponential rate 
of growth $G_n$ is maximized for 
$$
\gamma^*(p)=p-q= 2p-1\ .
$$
\end{theorem}

The argument is straightforward. Observe that the expected value is
$$
\EE (G_n)= \EE (G_1)=p\log (1+\gamma )+q \log (1-\gamma)=g(\gamma ).
$$
This function of the variable $\gamma$ has a derivative,
$$
g'(\gamma )=\frac{p}{1+\gamma} - \frac{q}{1-\gamma}=\frac{(2p-1)-\gamma}{1-\gamma^2} \ ,
$$
and $g(\gamma)\to 0$ when $\gamma\to 0^+$, and $g(\gamma) \to - \infty$ when $\gamma\to 1^-$. Also
$g'(\gamma) >0$ near $0$, and $g'$ is decreasing, so $g$ is concave.
Thus $g$ increases from $0$ to its maximum attained at 
$$
\gamma^*=\gamma^*(p)= 2p-1
$$
and then decreases to $-\infty$. 

\medskip

A corollary of this analysis is:

\begin{theorem} \textbf{(Kelly ruin threshold)}
Under the same conditions as before, there is a threshold value $\gamma_0(p) >\gamma^*(p)$, that is the only solution of the equation
$$
p\log (1+\gamma_0) +(1-p) \log (1 - \gamma_0) =0 \ ,
$$
such that for 
$\gamma >\gamma_0(p)$ we have almost sure ruin.
\end{theorem}

Indeed the equation is $g(\gamma_0) =0$ and since $g$ is decreasing for $\gamma >\gamma^*$ we have that for $\gamma > \gamma_0$, $g(\gamma) <0$, and 
the expected exponential growth is negative. 

\medskip

Behavioural considerations of Keynes ``animal spirits'', as discussed in the preliminaries, indicate that most players take more risks than allowed by the Kelly criterion and that the average 
$\gamma$ is larger than $\gamma_0$, which completes our argument.

\end{document}